\documentclass[11pt,a4paper]{article}

\usepackage[a4paper,margin=0.72in]{geometry}
\usepackage{graphicx}
\usepackage{amsmath,amssymb}
\usepackage{url}
\usepackage{xurl}
\usepackage{cite}
\usepackage{float}
\usepackage{placeins}
\usepackage{xcolor}
\usepackage{dsfont}
\usepackage{ragged2e}
\usepackage[colorlinks=true,linkcolor=blue,citecolor=blue,urlcolor=blue]{hyperref}
\usepackage[font=small,labelfont=bf,justification=centering]{caption}

\begin{document}

\renewcommand{\thefootnote}{\fnsymbol{footnote}}

\twocolumn[
\begin{center}
{\Large\bfseries Collective drift and pinning in active rotator networks with Kuramoto coupling and mixed-sign feedback disorder\par}
\vspace{0.6em}
Arpan Dey $^{1,*}$\\
$^1$ Universit\'e de Montpellier, Montpellier, France\\
Email: \texttt{arpand2004@gmail.com}
\end{center}

\vspace{0.4em}
\begin{center}
\begin{minipage}{0.82\textwidth}
\small
\justifying
Active rotator models provide a minimal phase description of excitable and oscillatory systems and have long been used to study mutual entrainment, synchronization, and collective transitions. Here we investigate fully connected active rotator networks with Kuramoto coupling, where a common positive intrinsic drive competes with local feedback amplitudes drawn from a zero-mean Gaussian distribution. We characterize the dynamics using the mean absolute late-time drift, the polar and nematic order parameters, and the fractions of positive and negative drifting oscillators. The model with an even mixed-sign feedback ensemble has an exact discrete symmetry under a simultaneous phase shift by $\pi$ and reversal of the feedback amplitudes. This exposes two physically distinct low-drift states: at weak coupling and strong disorder, local pinning produces an approximately antipodal, nematically ordered population, whereas at strong coupling the system can form a polar, coherently pinned state. A self-consistent stationary mean-field theory determines the existence edge of a nonzero coherent branch, provides quantitative predictions for the polar and nematic order, and shows that the coherent branch followed here is fully pinned. Sustained negative drift is impossible in a stationary mean field; the negative-drift population observed is instead localized to an intermediate-coupling regime characterized by a non-stationary mean field. Finally, finite-size extrapolations show persistent coupling-induced drift recovery and are compatible with vanishing strong-coupling drift at large disorder in the thermodynamic limit. These results show that even with a homogeneous intrinsic drive, mixed-sign local feedback can generate distinct regimes of drift, pinning, and collective phase organization.\end{minipage}
\end{center}
\vspace{0.8em}
]

\footnotetext[1]{Corresponding author.}
\renewcommand{\thefootnote}{\arabic{footnote}}

\section{Introduction}

Synchronization in populations of coupled oscillators is a central problem in nonlinear dynamics, with applications ranging from chemical and biological rhythms to condensed-matter and networked systems. The Kuramoto model provides one of the simplest and most successful descriptions of collective phase synchronization, showing how mutual coupling can overcome intrinsic heterogeneity and produce macroscopic coherence \cite{acebron2005}. Active rotator models extend this phase-oscillator picture by including an additional local term that can make an individual unit either oscillatory or excitable, depending on the balance between intrinsic drive and local pinning. They therefore provide a natural framework for studying the competition between individual pinning and collective entrainment.

The classical active rotator literature has shown that coupled excitable or oscillatory elements can undergo collective transitions, including the emergence of macroscopic rhythms and hysteretic behavior in mean-field settings \cite{shinomoto1986,sakaguchi1986,sakaguchi1988}. Related work on externally driven coupled oscillator systems, including systems with random pinning and external drive, has shown how pinned states, depinning transitions, coherent moving states, and cooperative oscillator dynamics can arise as coupling or drive strength is varied \cite{strogatz1989}. More recently, random-field Kuramoto models have examined the effect of random pinning fields on synchronization, both on complete graphs and on random complex networks \cite{lopes2016}. These studies show that local pinning fields, external drive, and collective phase alignment can interact in subtle ways.

In this work, we study a closely related but distinct problem. We consider fully connected active rotator networks with Kuramoto coupling and a common intrinsic drive, while each local feedback amplitude is drawn from a zero-mean mixed-sign distribution. Thus all oscillators are driven in the same intrinsic direction, but their local feedback terms can either support or oppose the drive depending on the sign of the local feedback amplitude. This makes the feedback disorder different from a conventional spread of intrinsic frequencies: the randomness enters through the local pinning mechanism rather than through the intrinsic drive itself. Increasing the feedback disorder strength tends to suppress drift locally. Strong coupling then has two distinct outcomes: at weak or moderate disorder it can reorganize the population into a coherent drifting state, whereas at sufficiently large disorder it can instead organize the population into a coherent polar pinned state.

We analyze this competition using numerical regime maps in the feedback-disorder--coupling plane. In addition to the mean absolute late-time drift, we use the usual first-harmonic Kuramoto order parameter and a second-harmonic nematic order parameter. Their distinction is essential because the exact $\pi$-shift symmetry of the disorder ensemble permits strong phase organization with little polar order. Unlike the continuous global phase-shift invariance of the standard Kuramoto interaction, the local feedback term leaves only a discrete statistical pairing of symmetry-related collective states. We derive exact uncoupled benchmarks, a necessary condition for collective pinning, a self-consistent stationary mean-field theory, and an asymptotic coherent-phase scaling estimate. The stationary theory is tested directly against simulation in the coherent-pinning regime, showing quantitative agreement for the polar order and capturing the nematic order to within a few hundredths. We then resolve positive and negative drift separately, identify the origin of backward slips, and examine the thermodynamic limit using finite-size extrapolations.

\section{Model and analytical results}

\subsection{Model, observables, and symmetry}
\label{sec:model-observables}

We consider a fully connected network of $N$ active rotators with a common intrinsic drive, mixed-sign feedback disorder, and Kuramoto coupling. The phase of oscillator $i$ evolves as

\begin{equation}\label{eq:model-dimensional}
\frac{d\theta_i}{dt}=\omega-A_i\sin\theta_i+\frac{K}{N-1}\sum_{j\ne i}\sin(\theta_j-\theta_i).
\end{equation}

Here $\omega>0$ is fixed and identical for all oscillators, $K$ is the global coupling strength, and $A_i$ is the local feedback amplitude, sampled from a zero-mean Gaussian distribution with standard deviation $\sigma_A$,

\begin{equation}\label{eq:feedback-distribution-dimensional}
A_i\sim\mathcal{N}(0,\sigma_A^2).
\end{equation}

The coupling term tends to align oscillator phases, while the local feedback produces oscillator-dependent, phase-dependent forcing that can accelerate or decelerate the motion and, when sufficiently strong, pin an oscillator. Since $A_i$ can have either sign, the preferred phase of the local feedback is reversed between positive- and negative-feedback oscillators.

We introduce the dimensionless variables

\begin{equation}\label{eq:dimensionless-variables}
\tau=\omega t,\qquad a_i=\frac{A_i}{\omega},\qquad \kappa=\frac{K}{\omega},\qquad r=\frac{\sigma_A}{\omega}.
\end{equation}

Equation~\eqref{eq:model-dimensional} becomes

\begin{equation}\label{eq:dimless}
\frac{d\theta_i}{d\tau}=1-a_i\sin\theta_i+\frac{\kappa}{N-1}\sum_{j\ne i}\sin(\theta_j-\theta_i),
\end{equation}

with $a_i\sim\mathcal{N}(0,r^2)$.

To characterize phase organization, we define the first two circular harmonics,

\begin{equation}\label{eq:order-parameters}
R_m e^{\mathrm{i}\psi_m}=\frac{1}{N}\sum_{j=1}^{N}e^{\mathrm{i}m\theta_j},\qquad m=1,2.
\end{equation}

where $\mathrm{i}$ is the imaginary unit. $R_1$ is the usual polar or Kuramoto coherence: $R_1\simeq1$ indicates a narrow single phase cluster. $R_2$ measures nematic coherence: it is insensitive to a shift by $\pi$, so two narrow clusters separated by $\pi$ can have $R_2\simeq1$ while $R_1\simeq0$. The coupling depends only on the first harmonic,

\begin{equation}\label{eq:coupling-order-parameter}
\frac{\kappa}{N-1}\sum_{j\ne i}\sin(\theta_j-\theta_i)=\kappa\frac{N}{N-1}R_1\sin(\psi_1-\theta_i).
\end{equation}

Thus, for large $N$,

\begin{equation}\label{eq:largeN-model}
\frac{d\theta_i}{d\tau}=1-a_i\sin\theta_i+\kappa R_1\sin(\psi_1-\theta_i).
\end{equation}

The principal dynamical observable is the mean absolute late-time drift,

\begin{equation}\label{eq:drift-observable}
D(r,\kappa)=\left\langle\frac{1}{N}\sum_{i=1}^{N}\frac{|\Omega_i|}{\omega}\right\rangle,
\end{equation}

where $\Omega_i$ is the late-time average angular velocity of oscillator $i$ and the outer brackets denote averaging over numerical trials. The absolute value prevents opposite drift directions from cancelling. In the uncoupled limit we write $D_0(r)=D(r,0)$.

To resolve the direction of motion, we define the fractions of positively and negatively drifting oscillators using the normalized late-time velocity,

\begin{equation}\label{eq:fplus}
f_+=\left\langle\frac{1}{N}\sum_{i=1}^{N}\mathds{1}\!\left(\frac{\Omega_i}{\omega}>\epsilon\right)\right\rangle,
\end{equation}

and

\begin{equation}\label{eq:fminus}
f_-=\left\langle\frac{1}{N}\sum_{i=1}^{N}\mathds{1}\!\left(\frac{\Omega_i}{\omega}<-\epsilon\right)\right\rangle,
\end{equation}

with $\epsilon=10^{-3}$. Oscillators with $|\Omega_i|/\omega\le\epsilon$ are counted as effectively pinned.

The mixed-sign model has an exact discrete transformation. If $\{\theta_i(\tau)\}$ solves Eq.~\eqref{eq:dimless} for feedback amplitudes $\{a_i\}$, then

\begin{equation}\label{eq:pi-symmetry}
\theta_i(\tau)\longrightarrow\theta_i(\tau)+\pi,\qquad a_i\longrightarrow-a_i
\end{equation}

produces another solution. Since the Gaussian distribution is symmetric, $g_r(a)=g_r(-a)$, the disorder ensemble is invariant under Eq.~\eqref{eq:pi-symmetry}; the uniform distribution of initial phases used in the simulations is likewise invariant under a global shift by $\pi$. Consequently, when both symmetry-related branches are sampled with equal statistical weight, the ensemble-averaged complex first harmonic vanishes, whereas the second harmonic is unchanged.

This symmetry differs from the continuous global phase-shift invariance of the standard Kuramoto model with purely phase-difference coupling. Here the term $-a_i\sin\theta_i$ anchors the phase and removes that continuous degeneracy.

\subsection{Uncoupled limit: local pinning, drift, and nematic order}

Before studying the coupled system, it is useful to examine the local dynamics of a single oscillator. Setting $K=0$ and dropping the oscillator index gives

\begin{equation}\label{eq:uncoupled-single}
\frac{d\theta}{d\tau}=1-a\sin\theta.
\end{equation}

A fixed point satisfies

\begin{equation}\label{eq:uncoupled-fixed-point}
\sin\theta^*=\frac{1}{a},
\end{equation}

and therefore exists only when

\begin{equation}\label{eq:local-pinning-condition}
|a|\ge1.
\end{equation}

\begin{figure*}[!t]
\centering
\includegraphics[width=0.82\textwidth,height=0.76\textheight,keepaspectratio]{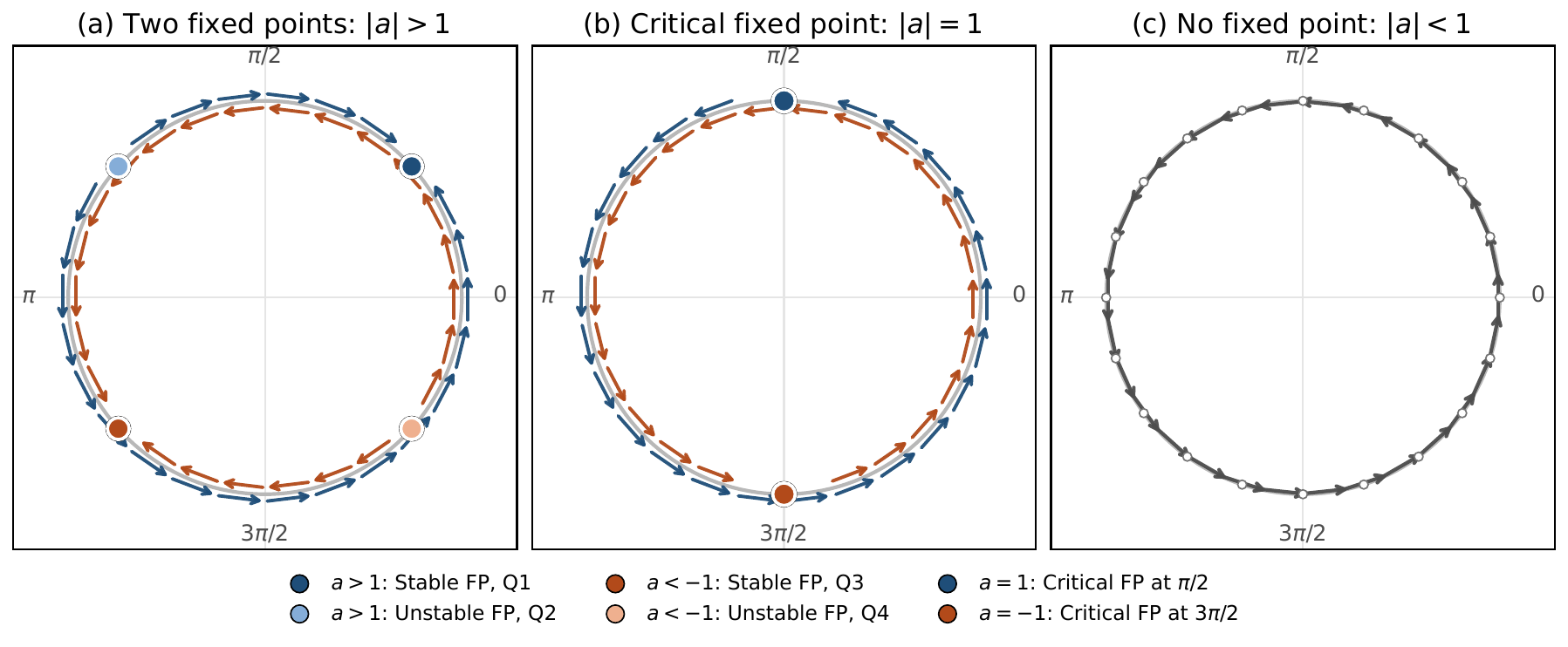}
\caption{Phase-circle schematics for the local dynamics $\dot{\theta}=\omega(1-a\sin\theta)$. (a) Two-fixed-point regime $|a|>1$. (b) Critical case $|a|=1$. (c) Drifting regime $|a|<1$. Stable, unstable, and critical fixed points are marked for the corresponding positive and negative values of $a$.}
\label{fig:local-dynamics}
\end{figure*}

For $|a|>1$ there is one stable and one unstable fixed point. For $a>1$ the stable point lies in the first quadrant; for $a<-1$ it lies in the third quadrant. At $|a|=1$ the two fixed points merge into a critical fixed point, while for $|a|<1$ the oscillator drifts continuously around the phase circle. These regimes are summarized in Fig.~\ref{fig:local-dynamics}.

\begin{figure*}[!t]
\centering
\includegraphics[width=0.82\textwidth,height=0.78\textheight,keepaspectratio]{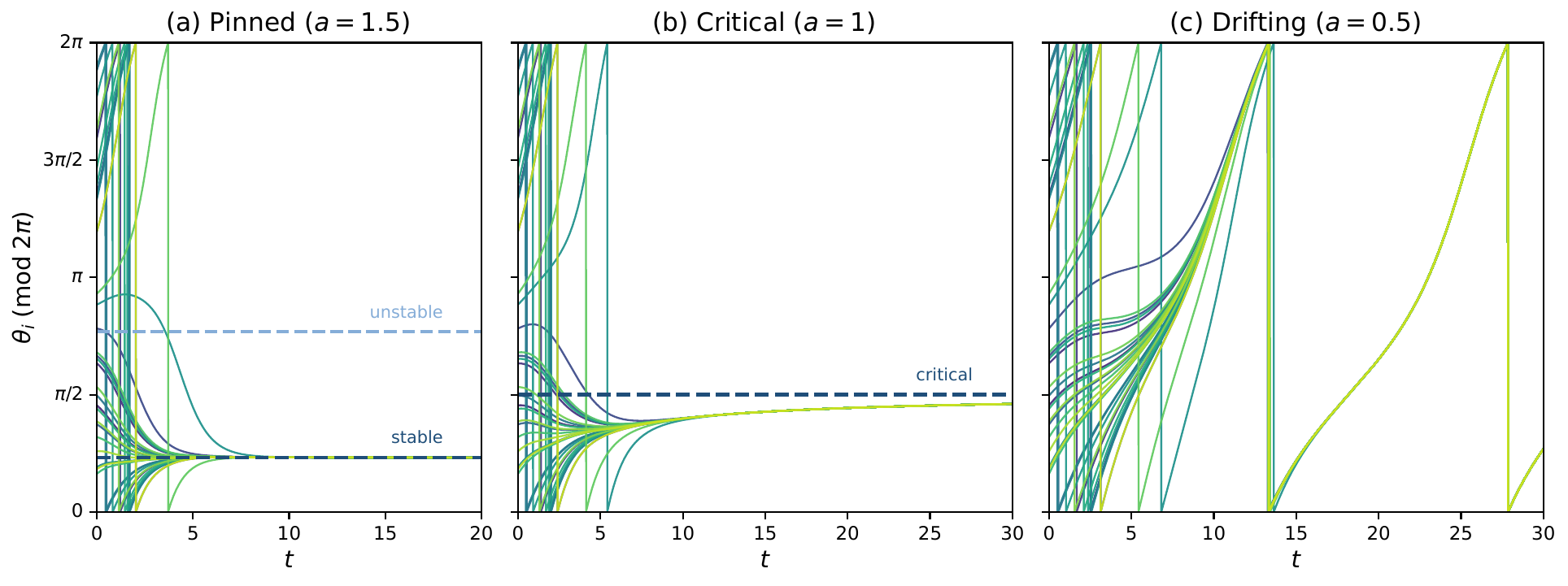}
\caption{Wrapped phase trajectories for an illustrative homogeneous population of $N=30$ fully connected oscillators, with the same dimensionless feedback amplitude $a=A/\omega$ assigned to every oscillator. Phases are initialized uniformly in $[0,2\pi)$ and integrated by the Euler method with $\omega=0.5$ and $K=0.6$. Panels (a)--(c) show $a=1.5$, $a=1$, and $a=0.5$, respectively, corresponding to pinned, critical, and drifting local regimes.}
\label{fig:wrapped-trajectories}
\end{figure*}

This local classification remains useful after restoring the Kuramoto coupling. Figure~\ref{fig:wrapped-trajectories} shows illustrative wrapped phase trajectories for a fully connected homogeneous population in which all oscillators are assigned the same dimensionless feedback amplitude $a=A/\omega$. For $a=1.5$ the oscillators approach the stable fixed point, for $a=1$ they slow near the critical fixed point, and for $a=0.5$ they continue to drift. These examples isolate the local pinning mechanism before the mixed-sign disorder of the main simulations is restored.

For $|a|<1$, the period of one full rotation is

\begin{equation}\label{eq:uncoupled-period-integral}
P(a)=\int_0^{2\pi}\frac{d\theta}{1-a\sin\theta},
\end{equation}

which gives

\begin{equation}\label{eq:uncoupled-period}
P(a)=\frac{2\pi}{\sqrt{1-a^2}}.
\end{equation}

The normalized late-time drift of a drifting oscillator is therefore

\begin{equation}\label{eq:single-oscillator-drift}
\frac{\Omega(a)}{\omega}=\frac{2\pi}{P(a)}=\sqrt{1-a^2},\qquad |a|<1.
\end{equation}

The dimensionless feedback density is

\begin{equation}\label{eq:gaussian-density}
g_r(a)=\frac{1}{\sqrt{2\pi r^2}}\exp\left(-\frac{a^2}{2r^2}\right).
\end{equation}

Averaging Eq.~\eqref{eq:single-oscillator-drift} over the drifting part of the distribution gives the exact uncoupled benchmark

\begin{equation}\label{eq:D0-integral}
D_0(r)=\int_{-1}^{1}\sqrt{1-a^2}\,g_r(a)\,da.
\end{equation}

For weak disorder, using $\langle a^2 \rangle = r^2$,

\begin{equation}\label{eq:D0-small-r}
D_0(r)\simeq1-\frac{r^2}{2},\qquad r\ll1,
\end{equation}

while for broad feedback disorder,

\begin{equation}\label{eq:D0-large-r}
D_0(r)\simeq\frac{\sqrt{\pi/8}}{r},\qquad r\gg1.
\end{equation}

Thus the uncoupled drift decays only as $1/r$ at large $r$: at any finite $r$ there remains a small population with $|a|<1$ that continues to drift.

The same uncoupled problem admits an exact second-harmonic benchmark. Let $m_2(a)$ denote the long-time second-harmonic moment $\langle e^{2\mathrm{i}\theta}\rangle$ of a single oscillator with feedback amplitude $a$. For $|a|<1$, averaging over the invariant phase density $p(\theta|a)\propto[1-a\sin\theta]^{-1}$ gives

\begin{equation}\label{eq:m2-drift}
m_2^{\mathrm{drift}}(a)
=1-\frac{2}{a^2}\left(1-\sqrt{1-a^2}\right),
\qquad |a|<1,
\end{equation}

with the removable limit $m_2^{\mathrm{drift}}(0)=0$. For $|a|\ge1$, the oscillator approaches a stable fixed point $\theta^*$. Evaluating $e^{2\mathrm{i}\theta^*}$ then gives

\begin{equation}\label{eq:m2-pinned}
m_2^{\mathrm{pinned}}(a)
=1-\frac{2}{a^2}
+\mathrm{i}\,\frac{2}{|a|}\sqrt{1-a^{-2}},
\qquad |a|\ge1.
\end{equation}

Averaging this single-oscillator response over the feedback distribution gives the uncoupled nematic order,

\begin{equation}\label{eq:R20}
R_2^0(r)
=\left|\int_{-\infty}^{\infty}g_r(a)\,m_2(a)\,da\right|,
\end{equation}

where $m_2=m_2^{\mathrm{drift}}$ for $|a|<1$ and $m_2=m_2^{\mathrm{pinned}}$ otherwise. The two branches match continuously at $|a|=1$. At $r=0$ the phase distribution remains uniform and $R_2^0(0)=0$; as $r\to\infty$, almost all oscillators pin near $0$ or $\pi$, so $R_2^0(r)\to1$ even though the ensemble-averaged polar order vanishes by symmetry.

\subsection{Necessary condition for collective pinning}

We next derive a simple necessary condition for collective pinning in the fully connected system. Averaging Eq.~\eqref{eq:dimless} over all oscillators removes the pairwise coupling contribution by antisymmetry

\begin{equation}\label{eq:population-velocity}
\left\langle\frac{d\theta_i}{d\tau}\right\rangle_i=1-\langle a_i\sin\theta_i\rangle_i.
\end{equation}

A collectively pinned state must have zero population-averaged velocity, and therefore

\begin{equation}\label{eq:collective-pinning-balance}
1=\langle a_i\sin\theta_i\rangle_i.
\end{equation}

This is not sufficient for every oscillator to be individually pinned, but it shows that collective pinning requires a nontrivial correlation between feedback amplitudes and phases. Since $|\sin\theta_i|\le1$,

\begin{equation}\label{eq:necessary-abs-feedback}
\langle|a_i|\rangle_i\ge1.
\end{equation}

For $a_i\sim\mathcal{N}(0,r^2)$, the population average converges to $\langle|a_i|\rangle_i=r\sqrt{2/\pi}$ in the thermodynamic limit, so

\begin{equation}\label{eq:necessary-r}
r\ge\sqrt{\frac{\pi}{2}}\simeq1.253.
\end{equation}

Below Eq.~\eqref{eq:necessary-r}, the available local feedback disorder is too weak on average to balance the common drive. Above it, collective pinning is allowed but not guaranteed: the phases must still organize so that Eq.~\eqref{eq:collective-pinning-balance} is satisfied.

\subsection{Effective feedback amplitude in a stationary mean field}

Equation~\eqref{eq:largeN-model} can be rewritten exactly by combining the local feedback and the collective field. For a stationary first-harmonic mean field, let us define

\begin{equation}\label{eq:complex-mean-field}
Z\equiv\kappa R_1 e^{\mathrm{i}\psi_1}.
\end{equation}

For an oscillator with feedback amplitude $a$, we write

\begin{equation}\label{eq:effective-field}
\begin{aligned}
a+Z&=B(a)e^{\mathrm{i}\delta(a)},\\
B(a)&=|a+Z|,\qquad \delta(a)=\arg(a+Z).
\end{aligned}
\end{equation}

Then Eq.~\eqref{eq:largeN-model} becomes

\begin{equation}\label{eq:effective-adler}
\frac{d\theta}{d\tau}=1-B(a)\sin[\theta-\delta(a)].
\end{equation}

When $R_1$ and $\psi_1$ are constant, $B$ and $\delta$ are constant as well. With $u=\theta-\delta$, Eq.~\eqref{eq:effective-adler} is the same equation as Eq.~\eqref{eq:uncoupled-single}. Its late-time average angular velocity is therefore

\begin{equation}\label{eq:stationary-rotation-number}
\frac{\Omega(a)}{\omega}=\begin{cases}
\sqrt{1-B(a)^2},&B(a)<1,\\[3pt]
0,&B(a)\ge1.
\end{cases}
\end{equation}

Thus, in a stationary mean field,

\begin{equation}\label{eq:no-negative-stationary}
\Omega(a)\ge0
\end{equation}

for every oscillator. Sustained negative late-time drift is therefore impossible under a stationary mean field, irrespective of $a$, $r$, or $\kappa$. This result will be used below to identify the origin of the negative-drift population observed in simulations.

\subsection{Self-consistent stationary mean-field theory}

We now close the stationary mean-field problem in the thermodynamic limit. For a given $B$, the long-time first-harmonic response can be written as

\begin{equation}\label{eq:h1B}
h_1(B)=\begin{cases}
\sqrt{1-B^{-2}}+\mathrm{i}/B,&B\ge1,\\[3pt]
\dfrac{\mathrm{i}}{B}\left(1-\sqrt{1-B^2}\right),&B<1,
\end{cases}
\end{equation}

with $h_1(0)=0$ understood by continuity.

Since $e^{\mathrm{i}\delta(a)}=(a+Z)/|a+Z|$, the long-time contribution of oscillators with feedback amplitude $a$ is

\begin{equation}\label{eq:single-mean-field-response}
\left\langle e^{\mathrm{i}\theta}\right\rangle_a=\frac{a+Z}{|a+Z|}\,h_1(|a+Z|).
\end{equation}

At the isolated point $a+Z=0$, the right-hand side of Eq.~\eqref{eq:single-mean-field-response} is defined by its continuous limit, which is zero. 

Define the disorder-averaged first-harmonic response

\begin{equation}\label{eq:disorder-response}
I(Z;r)\equiv\int_{-\infty}^{\infty}g_r(a)\,
\frac{a+Z}{|a+Z|}\,h_1(|a+Z|)\,da.
\end{equation}

From the earlier definition $Z=\kappa R_1e^{\mathrm{i}\psi_1}$, the mean field must equal $\kappa$ times the population-averaged first harmonic. Therefore a stationary solution must satisfy the self-consistency condition

\begin{equation}\label{eq:self-consistency}
Z=\kappa I(Z;r).
\end{equation}

 For the drifting sector $B<1$, Eq.~\eqref{eq:h1B} is the time average with respect to the invariant density, $p(u|a)\propto[1-B(a)\sin u]^{-1}$. The stationary closure therefore assumes that the drifting subpopulation is dephased and samples this invariant density rather than remaining coherently bunched. For the continuum of Gaussian feedback amplitudes, the $a$-dependence of the angular velocity provides the usual dephasing mechanism.

For any nonzero solution at $\kappa>0$,

\begin{equation}\label{eq:R1-psi-from-Z}
R_1=\frac{|Z|}{\kappa},\qquad \psi_1=\arg Z.
\end{equation}

The same solution predicts the mean absolute drift,

\begin{equation}\label{eq:MF-drift}
D_{\mathrm{MF}}(r,\kappa)=\int_{|a+Z|<1}\sqrt{1-|a+Z|^2}\,g_r(a)\,da,
\end{equation}

and the pinned fraction,

\begin{equation}\label{eq:MF-pinned-fraction}
f_{\mathrm{pinned}}=\int_{|a+Z|\ge1}g_r(a)\,da.
\end{equation}

A simple sufficient condition for complete pinning follows from the imaginary part of $Z$. Since

\[
B(a)^2=|a+Z|^2=(a+\operatorname{Re}Z)^2+(\operatorname{Im}Z)^2,
\]

one has $B(a)\ge|\operatorname{Im}Z|$ for every real $a$. Therefore

\begin{equation}\label{eq:full-pinning-criterion}
|\operatorname{Im}Z|\ge1
\quad\Longrightarrow\quad
D_{\mathrm{MF}}=0,\qquad f_{\mathrm{pinned}}=1.
\end{equation}

The second harmonic follows from the corresponding single-oscillator response

\begin{equation}\label{eq:h2B}
h_2(B)=\begin{cases}
1-\dfrac{2}{B^2}+\dfrac{2\mathrm{i}}{B}\sqrt{1-B^{-2}},&B\ge1,\\[6pt]
1-\dfrac{2}{B^2}\left(1-\sqrt{1-B^2}\right),&B<1,
\end{cases}
\end{equation}

with $h_2(0)=0$. Hence

\begin{equation}\label{eq:MF-R2}
R_2=\left|\int_{-\infty}^{\infty}g_r(a)
\left(\frac{a+Z}{|a+Z|}\right)^2 h_2(|a+Z|)\,da\right|,
\end{equation}

where the integrand at $a+Z=0$ is again understood by continuity. 

At $Z=0$, the effective amplitude reduces to $B=|a|$. Comparing Eqs.~\eqref{eq:h2B}, \eqref{eq:m2-drift}, and \eqref{eq:m2-pinned} then gives
$h_2(|a|)=m_2(a)$. So Eq.~\eqref{eq:MF-R2} reduces to the uncoupled expression Eq.~\eqref{eq:R20} when $Z=0$. Likewise Eq.~\eqref{eq:MF-drift} reduces to Eq.~\eqref{eq:D0-integral}. The self-consistent construction therefore unifies the exact uncoupled benchmarks with the stationary coupled theory.

The symmetry of Eq.~\eqref{eq:pi-symmetry} appears directly in the response integral. Because $g_r(a)$ is even, a change of variable $a\to-a$ gives

\begin{equation}\label{eq:Z-pair-symmetry}
I(-Z;r)=-I(Z;r).
\end{equation}

In particular, Eq.~\eqref{eq:Z-pair-symmetry} gives $I(0;r)=0$, so $Z=0$ is always a self-consistent stationary solution with vanishing first-harmonic mean field; higher-harmonic order such as $R_2$ may nevertheless remain nonzero.

Thus, if a nonzero stationary solution $Z$ exists, the symmetry-related solution $-Z$ exists as well, with the same $R_1$ and a collective phase shifted by $\pi$. A nonzero $Z$ here is a branch-resolved mean field. It does not contradict the vanishing ensemble-averaged complex first harmonic discussed in Sec.~\ref{sec:model-observables}: the theoretical quantity $R_1=|Z|/\kappa$ is compared with the trial-averaged modulus measured numerically, whereas an equal-weight complex average over the two symmetry-related branches vanishes.

Equation~\eqref{eq:self-consistency} is nonlinear and can admit more than one stationary root. We evaluate the Gaussian integrals numerically with 140-point Gauss--Hermite quadrature. For the curve shown in Fig.~\ref{fig:drift-regime-map}, we specifically follow the high-coherence solution at $\kappa=6$ and continue that branch downward in $\kappa$. We denote by $\kappa_c^{\mathrm{MF}}(r)$ the existence edge of this continued coherent branch, i.e., the lowest coupling reached by the continuation; its stability and dynamical selection are not addressed here. The construction also assumes stationary $R_1$ and $\psi_1$. A rotating mean field cannot in general be made stationary by a simple change of frame because the local term $-a\sin\theta$ would then become explicitly time dependent.

For the coherent branch used to construct Fig.~\ref{fig:drift-regime-map}, the criterion in Eq.~\eqref{eq:full-pinning-criterion} is satisfied throughout the sampled continuation. In particular, every stored existence-edge point over $4.3\le r\le6$ has $|\operatorname{Im}Z|>1$; the smallest edge value is approximately $1.30$. The stationary mean-field prediction on this tracked branch is therefore exactly $D_{\mathrm{MF}}=0$ and $f_{\mathrm{pinned}}=1$.

To test the branch-resolved predictions beyond the location of the existence edge, Table~\ref{tab:mf-validation} compares the stationary theory with direct-simulation data at two coherent-pinning points. The simulation counterpart of $f_{\mathrm{pinned}}$ is the operational fraction $1-f_+-f_-$ defined with the same threshold $\epsilon=10^{-3}$ stated above.

\begin{table*}[!t]
\centering
\small
\renewcommand{\arraystretch}{1.15}
\begin{tabular}{lcccc}
\hline
& \multicolumn{2}{c}{$r=5,\ \kappa=6$} & \multicolumn{2}{c}{$r=6,\ \kappa=6$}\\
Quantity & Mean field & Simulation ($N=200$) & Mean field & Simulation ($N=100$)\\
\hline
$R_1$ & 0.7808 & $0.7807\pm0.0014$ & 0.7054 & $0.7165\pm0.0053$\\
$R_2$ & 0.4297 & $0.4050\pm0.0061$ & 0.3012 & $0.2870\pm0.0175$\\
$D$ & 0 & 0 & 0 & 0\\
$f_{\mathrm{pinned}}$ & 1 & 1 & 1 & 1\\
\hline
\end{tabular}
\caption{Stationary mean-field predictions compared with direct simulations at two coherent-pinning points. The $(r,\kappa)=(5,6)$ simulation uses the $N=200$, 250-trial data of Fig.~\ref{fig:dynamical-regimes}; its $R_1$ value is a trial average of the late-time mean, while its $R_2$ value is a trial average of the final-time value. The $(6,6)$ values of $R_1$, $R_2$, and $D$ use the standard $N=100$, 50-trial cuts of Fig.~\ref{fig:order-parameters}, while its pinned fraction follows from $f_+=f_-=0$ in Fig.~\ref{fig:drifting-fractions}. Quoted uncertainties are standard errors over trials.}
\label{tab:mf-validation}
\end{table*}

The agreement is quantitative for the polar order and captures the second-harmonic order to within a few hundredths, while the vanishing drift and unit pinned fraction agree exactly at these two sampled coherent-pinning points. Close to the existence edge, finite-$N$ dynamics can still show a small residual drift even though the stationary coherent branch itself is fully pinned. For example, at $r=6$ and $\kappa=4.5$ the $N=100$ simulation gives $D=0.0076\pm0.0045$, while Eq.~\eqref{eq:full-pinning-criterion} gives $D_{\mathrm{MF}}=0$ on the continued stationary branch. 

\subsection{Coherent strong-coupling scale}
\label{sec:coherent-scale}

A coherent strong-coupling expansion provides a complementary asymptotic scale, but it is not a phase boundary. Let

\begin{equation}\label{eq:coherent-decomposition}
\theta_i=\Theta+\phi_i,\qquad \langle\phi_i\rangle_i=0,
\end{equation}

where $\Theta$ is the coherent-cluster centroid. For $R_1\simeq1$, $\psi_1\simeq\Theta$, and $|\phi_i|\ll1$, the stationary equation contains a common $O(\kappa^{-1})$ displacement together with a disorder-dependent contribution. Defining $\Theta$ as the centroid absorbs the common displacement. Equivalently, in the thermodynamic limit that common term makes no contribution to $\langle a_i\phi_i\rangle_i$ because the zero-mean disorder has $\langle a_i\rangle_i\to0$. The centered deviation is therefore

\begin{equation}\label{eq:centered-phi}
\phi_i\simeq-\frac{a_i\sin\Theta}{\kappa}+O(\kappa^{-2}).
\end{equation}

Combining Eq.~\eqref{eq:centered-phi} with the collective pinning balance Eq.~\eqref{eq:collective-pinning-balance} gives

\begin{equation}\label{eq:coherent-scale}
\kappa\simeq-\frac{r^2}{2}\sin(2\Theta).
\end{equation}

Because $\kappa>0$, Eq.~\eqref{eq:coherent-scale} requires $\sin(2\Theta)<0$, placing the coherent pinned branches in the second or fourth quadrant; the two choices are related by Eq.~\eqref{eq:pi-symmetry}. 

Since $|\sin(2\Theta)|\le1$, the same calculation gives the bound

\begin{equation}\label{eq:coherent-upper-scale}
\kappa\lesssim\frac{r^2}{2}.
\end{equation}

The expansion requires $\kappa\gg r$ for the bulk of the Gaussian distribution, so Eqs.~\eqref{eq:coherent-scale} and \eqref{eq:coherent-upper-scale} are controlled only asymptotically at large $r$. They should therefore be read as a coherent strong-coupling scale and bound.

\section{Numerical results}

\begin{figure*}[!t]
\centering
\includegraphics[width=0.75\textwidth,height=0.82\textheight,keepaspectratio]{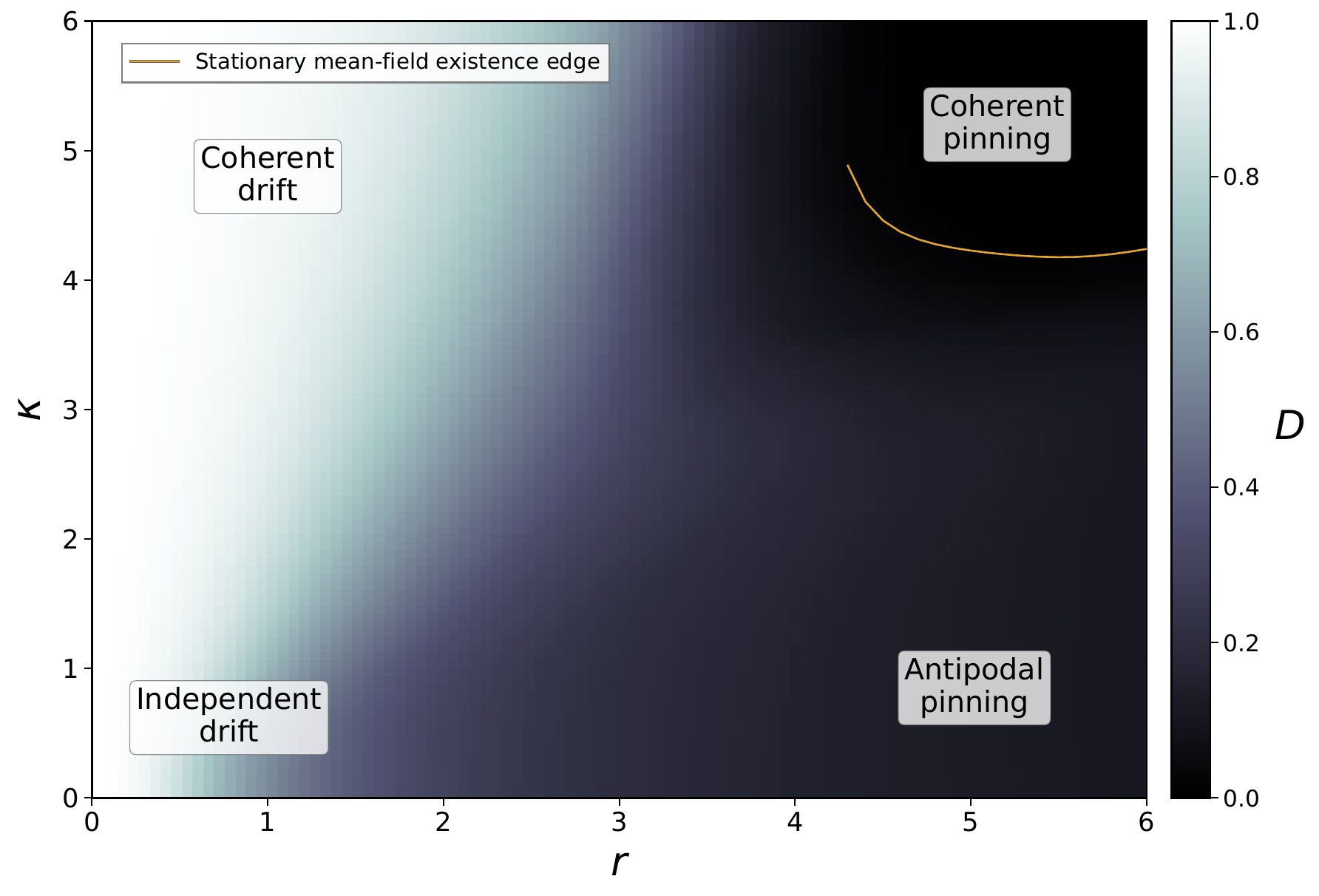}
\caption{Global drift landscape. Regime map of $D(r,\kappa)=\langle N^{-1}\sum_i|\Omega_i|/\omega\rangle$ for $N=100$, $\omega=0.5$, $dt=0.01$, $T=1000$, the final $30\%$ late-time window, and 50 trials. There is predominantly independent forward drift at low $r$ and low $\kappa$, coherent collective drift at low $r$ and high $\kappa$, antipodal local pinning with low drift at high $r$ and low $\kappa$, and coherent collective pinning at high $r$ and high $\kappa$. The solid curve is the stationary mean-field existence edge $\kappa_c^{\mathrm{MF}}(r)$ obtained by continuing the high-coherence stationary solution of Eq.~\eqref{eq:self-consistency} downward from $\kappa=6$.}
\label{fig:drift-regime-map}
\end{figure*}

\begin{figure*}[!t]
\centering
\includegraphics[width=0.85\textwidth,height=0.86\textheight,keepaspectratio]{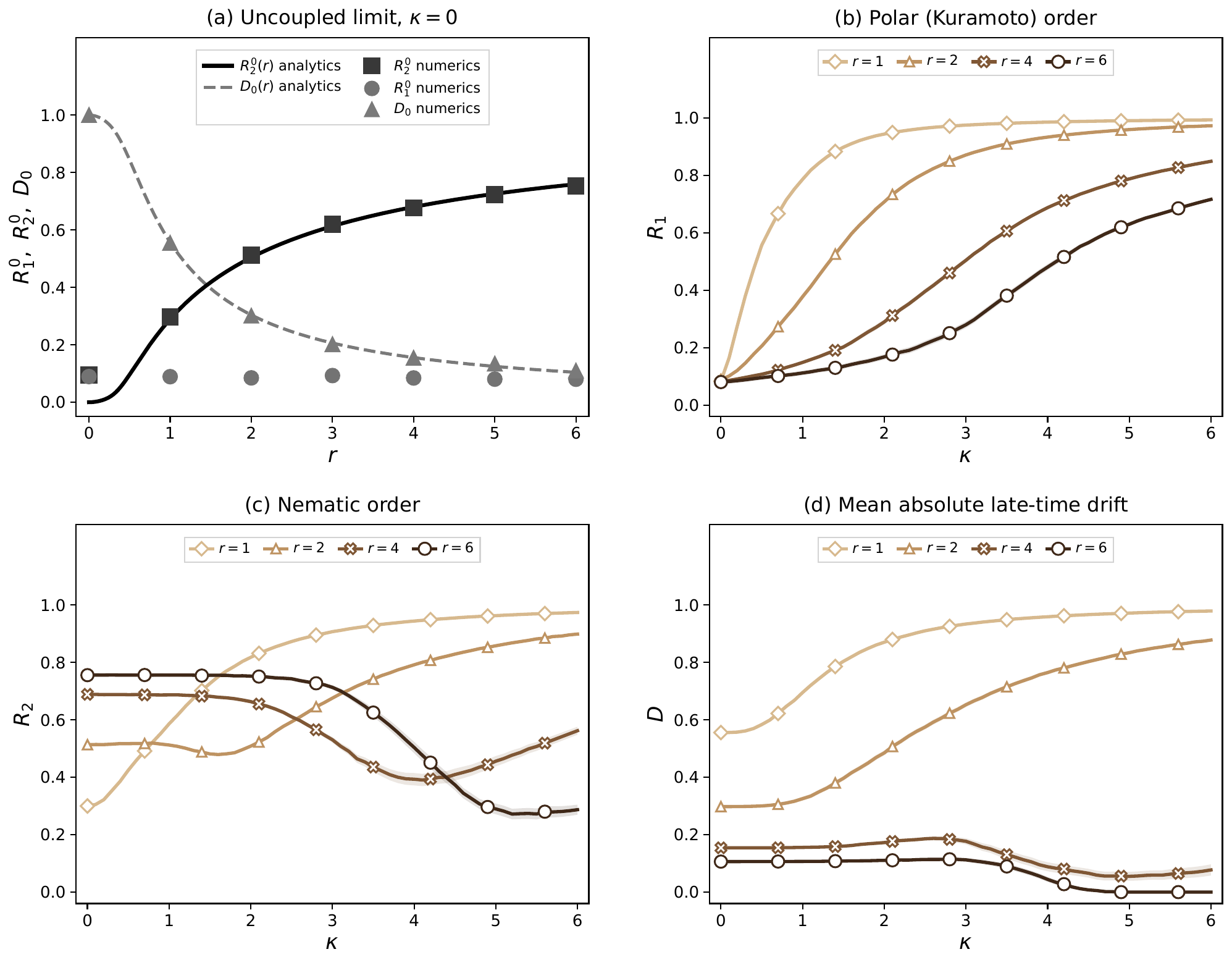}
\caption{Polar, nematic, and drift diagnostics. (a) Uncoupled limit $\kappa=0$: analytical $R_2^0(r)$ from Eq.~\eqref{eq:R20} and $D_0(r)$ from Eq.~\eqref{eq:D0-integral}, together with numerical $R_2^0$, $R_1^0$, and $D_0$. (b) Polar order $R_1$ versus $\kappa$ at fixed $r=1,2,4,6$. (c) Nematic order $R_2$ for the same cuts; at large $r$, $R_2$ decreases while $R_1$ increases, showing that the system evolves from an approximately antipodal phase organization toward a polar coherent state as coupling is increased. (d) Mean absolute late-time drift $D$ for the same cuts. Together, panels (b)--(d) distinguish coherent drift ($R_1,R_2,D$ large), antipodal low drift ($R_1$ small and $R_2$ large), and coherent pinning ($R_1$ large and $D\simeq0$).}
\label{fig:order-parameters}
\end{figure*}

Unless stated otherwise, simulations use $N=100$, $\omega=0.5$, Euler integration with $dt=0.01$ up to $T=1000$, uniformly random initial phases in $[0,2\pi)$, a late-time window comprising the final $30\%$ of each trajectory, and 50 independent trials. In Fig.~\ref{fig:order-parameters}, the reported $R_1$ and $R_2$ values are obtained by averaging the instantaneous order-parameter magnitudes over the late-time window and then averaging over trials. Figure~\ref{fig:dynamical-regimes} is a representative dynamical diagnostic; for this calculation, we use $N=200$ and 250 trials to obtain better-resolved data, while keeping the same $\omega$, $dt$, $T$, and late-time window. These data also provide the $(r,\kappa)=(5,6)$ simulation results used in Table~\ref{tab:mf-validation}; the $(6,6)$ column instead uses the standard $N=100$ simulations described above. The $N=200$ data of Fig.~\ref{fig:dynamical-regimes} do not enter the numerical regime map or thermodynamic-limit extrapolation; system-size dependence is assessed separately in Fig.~\ref{fig:finite-size}. Unless stated otherwise, error bars and shaded bands denote the standard error of the mean over trials.

\subsection{Global drift landscape}

Figure~\ref{fig:drift-regime-map} shows the regime map of the mean absolute late-time drift $D(r,\kappa)$. Increasing $r$ generally suppresses drift because a larger fraction of feedback amplitudes lie beyond the uncoupled local pinning threshold $|a|>1$. Coupling has a more nuanced role. At weak disorder and weak coupling the population largely consists of independently drifting rotators. Increasing $\kappa$ at weak or moderate disorder builds polar coherence and restores a fast collective forward drift. At large disorder and weak coupling the drift is again small, but for a different reason: local pinning produces an approximately antipodal phase population with strong nematic order and only residual independent drift. At large $r$ and large $\kappa$, the system instead forms a polar, coherently pinned state with large $R_1$ but $D\simeq0$.

The solid curve in Fig.~\ref{fig:drift-regime-map} is the stationary
mean-field existence edge $\kappa_c^{\mathrm{MF}}(r)$ obtained from
Eq.~\eqref{eq:self-consistency}. Over the large-$r$ part of the numerical
map, $4.5\lesssim r\le6$, the edge closely follows the onset of near-zero
drift: using the nearest sampled $r$ values, it lies within about $0.17$
in $\kappa$ of the first points with $D<0.02$. As shown by
Eq.~\eqref{eq:full-pinning-criterion}, the tracked stationary branch is
already fully pinned over this range; the smooth finite-$N$ decrease of
$D$ toward zero should therefore not be interpreted as a stationary
mean-field drifting branch or as evidence that the existence edge is a
stability boundary. 

\subsection{Polar and nematic organization}

Drift alone cannot determine how the phases are organized. Figure~\ref{fig:order-parameters} therefore compares the three complementary observables introduced in Sec.~\ref{sec:model-observables}: $D$ asks whether oscillators are moving, $R_1$ asks whether they are polar-aligned, and $R_2$ asks whether they are aligned modulo $\pi$.

Panel (a) provides two exact uncoupled benchmarks. The numerical $D_0$ follows Eq.~\eqref{eq:D0-integral}, and the numerical $R_2^0$ follows Eq.~\eqref{eq:R20}. At $r=0$, the continuum values of both $R_1^0$ and $R_2^0$ vanish, while the finite-$N$ simulations retain residual fluctuations. As $r$ increases, $D_0(r)$ decreases while $R_2^0(r)$ rises: disorder progressively converts free rotation into an approximately antipodal pinned population. In the large-$r$ limit, the two preferred sectors approach $0$ and $\pi$, so their first harmonics cancel in the symmetric ensemble while their second harmonics add.

Panels (b)--(d) show fixed-$r$ cuts as $\kappa$ is increased. At weak disorder, coupling produces a narrow moving cluster and $R_1$, $R_2$, and $D$ all approach unity. At larger $r$, $R_1$ rises more slowly because the coupling must reorganize a feedback-induced antipodal structure. The decrease of $R_2$ at intermediate $\kappa$ for $r=4$ and $r=6$, simultaneous with increasing $R_1$, resolves the reorganization from nematic toward polar order as coupling is increased. Most importantly, at strong disorder the high-coupling drift can approach zero while $R_1$ remains large. The large-$r$ weak-coupling and large-$r$ strong-coupling low-drift regions of Fig.~\ref{fig:drift-regime-map} are therefore physically distinct: the former is primarily antipodal and nematic, while the latter is a coherent polar pinned state.

\subsection{Directional drift and backward slips}

\begin{figure*}[!t]
\centering
\includegraphics[width=0.9\textwidth,height=0.82\textheight,keepaspectratio]{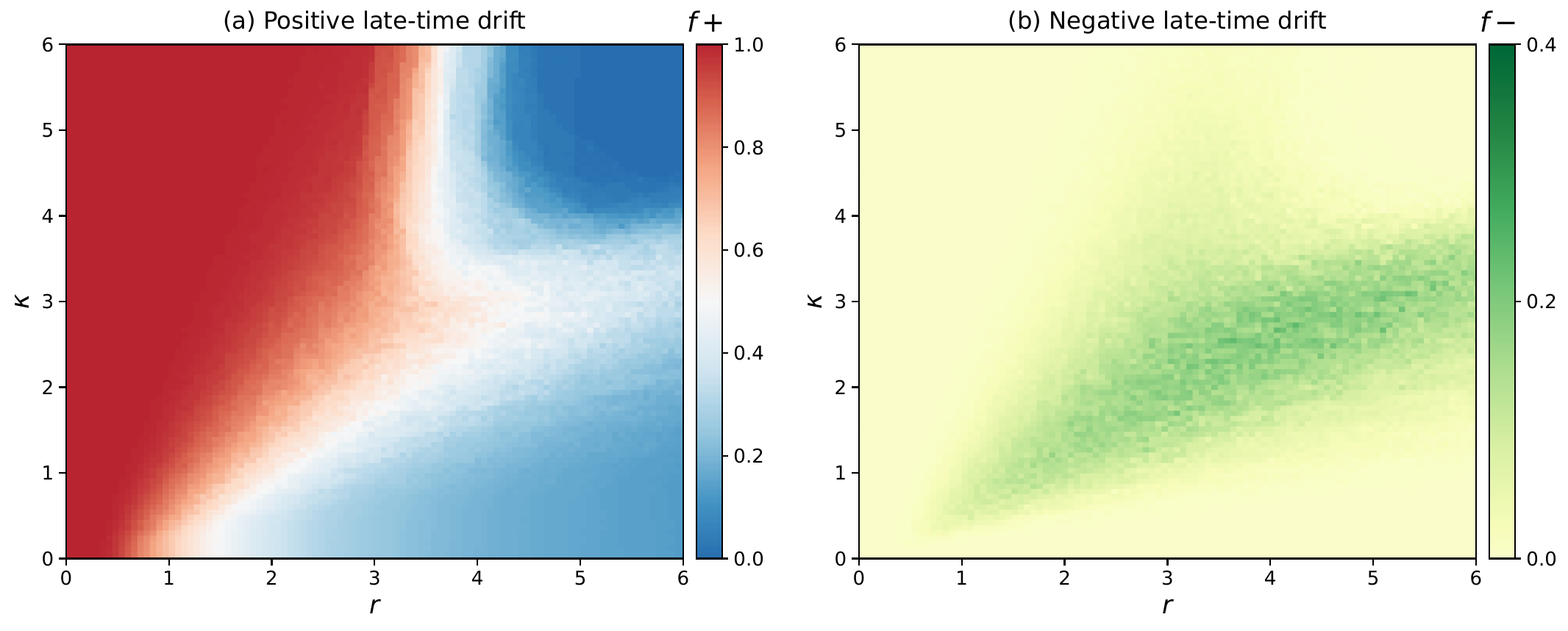}
\caption{Directional late-time drift fractions. (a) Positive fraction $f_+(r,\kappa)$ and (b) negative fraction $f_-(r,\kappa)$, defined by Eqs.~\eqref{eq:fplus} and \eqref{eq:fminus} with $\epsilon=10^{-3}$. Positive drift dominates the weak-disorder region, whereas negative drift is confined to an intermediate-coupling band and is absent along $\kappa=0$. At large $r$ and large $\kappa$, both fractions approach zero, identifying collective pinning rather than cancellation between oppositely drifting subpopulations.}
\label{fig:drifting-fractions}
\end{figure*}

\begin{figure*}[!t]
\centering
\includegraphics[width=1.0\textwidth,height=0.9\textheight,keepaspectratio]{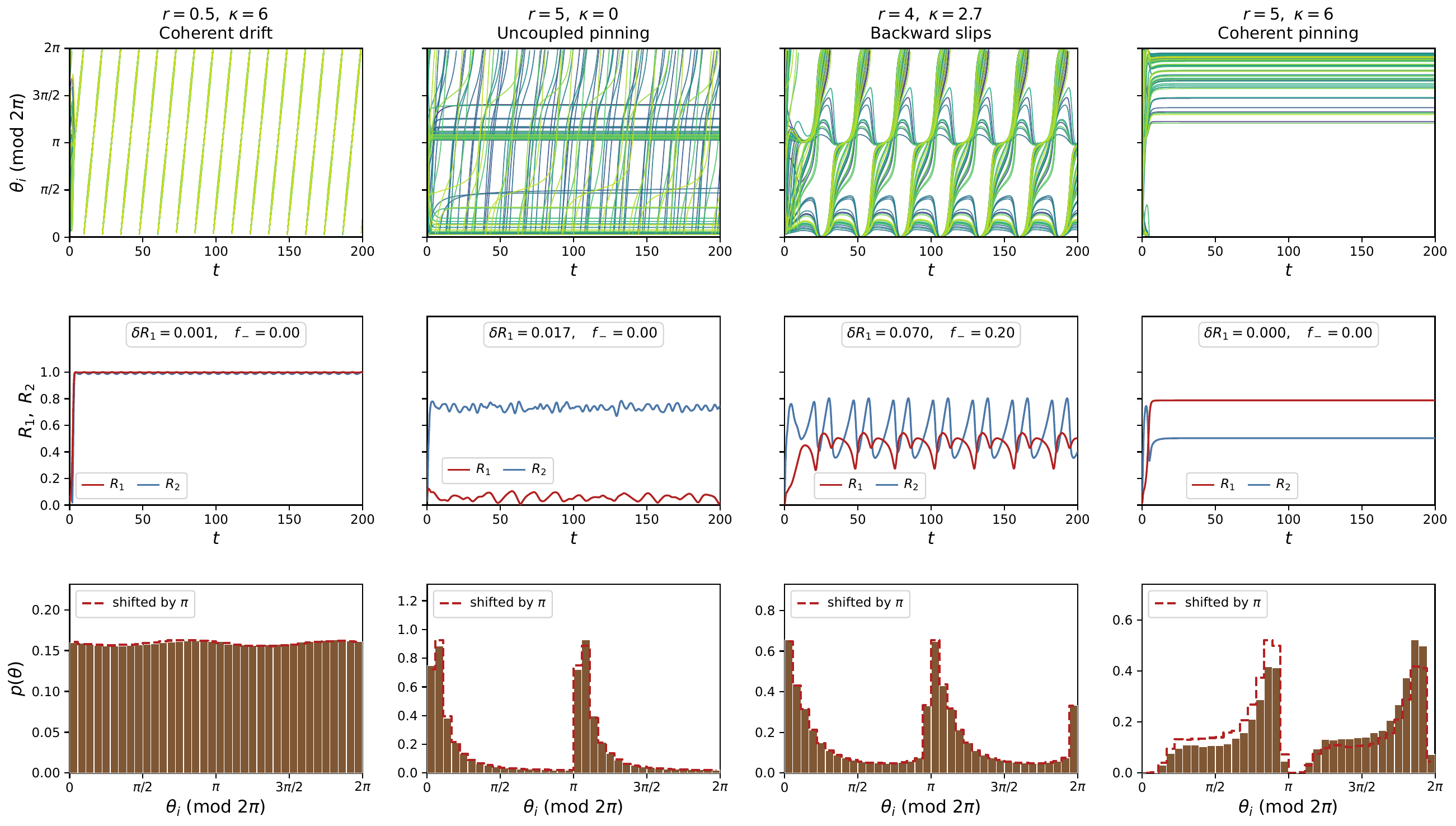}
\caption{Representative dynamics in four regimes for $N=200$: coherent drift $(r,\kappa)=(0.5,6)$, uncoupled local pinning $(5,0)$, backward slips $(4,2.7)$, and coherent collective pinning $(5,6)$. Top row: wrapped phase trajectories from one realization. Middle row: $R_1(t)$ and $R_2(t)$ from the same realization, with annotations showing $\delta R_1$ and $f_-$. Bottom row: late-time phase densities pooled over 250 trials; dashed curves show the same densities shifted by $\pi$. In the coherent-pinning column the overlap is imperfect because a finite ensemble does not sample the two symmetry-related branches with exactly equal weight. For readability, the time traces are displayed only up to $t=200$, whereas $\delta R_1$ and $f_-$ are computed from the full trajectories.}
\label{fig:dynamical-regimes}\end{figure*}

Figure~\ref{fig:drifting-fractions} shows $f_+$ and $f_-$. Positive drift dominates the weak-disorder region, especially at strong coupling where collective alignment reorganizes the phases into a coherent moving state and reduces the drift-suppressing effect of heterogeneous local feedback. The large-$r$, large-$\kappa$ corner is dark in both maps because the system is coherently pinned rather than drifting in either direction. Negative drift is much more localized: it appears mainly at intermediate coupling and moderate-to-large feedback disorder and vanishes identically along the entire $\kappa=0$ axis of the numerical map.

The stationary-field result of Eq.~\eqref{eq:no-negative-stationary} changes the interpretation of this region. A large local feedback amplitude cannot by itself create sustained backward rotation. For an uncoupled active rotator, $|a|>1$ produces pinning. A nonzero $f_-$ therefore requires a non-stationary mean field. If $R_1$ and $\psi_1$ vary in time, then $B_i$ and $\delta_i$ in Eq.~\eqref{eq:effective-field} also become time-dependent. With $u_i=\theta_i-\delta_i(\tau)$,

\begin{equation}\label{eq:time-dependent-adler}
\frac{du_i}{d\tau}=1-B_i(\tau)\sin u_i-\frac{d\delta_i}{d\tau},
\end{equation}

so the autonomous single-rotator classification no longer applies. 

Repeated excursions of the collective field can drive backward phase slips and thereby produce a negative late-time drift. The geometry of the effective field also gives, instantaneously,

\begin{equation}\label{eq:effective-field-bounds}
\bigl||a_i|-\kappa R_1(\tau)\bigr|
\le B_i(\tau)
\le |a_i|+\kappa R_1(\tau).
\end{equation}

Oscillators can repeatedly approach the pinning threshold only when the effective field can come sufficiently close to $B_i=1$. A necessary geometric condition is therefore

\begin{equation}\label{eq:marginal-band}
\bigl||a_i|-\kappa R_1(\tau)\bigr|\lesssim1.
\end{equation}

Since the typical feedback scale is $|a_i|\sim r$, Eq.~\eqref{eq:marginal-band} suggests competition between the local feedback and coupling fields on an approximately linear scale in the backward-slip regime. Extracting, for each $r$, the $\kappa$ value at which the numerical $f_-$ map is maximal gives a ridge that is well described over $1.5<r<5.5$ by

\begin{equation}\label{eq:negative-ridge}
\kappa_{\mathrm{ridge}}\simeq0.73\,r^{0.89},
\end{equation}

with a peak negative-drift fraction near $(r,\kappa)=(4.24,2.73)$.

To quantify the time dependence of the mean-field amplitude in representative runs, we use the late-time fluctuation

\begin{equation}
\label{eq:deltaR1}
\delta R_1=
\left\langle
\sqrt{{\langle R_1^2\rangle}_t-{\langle R_1\rangle}_t^2}
\right\rangle,
\end{equation}

where $\langle\cdot\rangle_t$ denotes the standard late-time window.

Figure~\ref{fig:dynamical-regimes} makes the mechanism visible in time. The four columns sample coherent drift, uncoupled local pinning, the backward-slip regime, and coherent collective pinning. In coherent drift, the oscillators form a narrow cluster that moves around the circle, so $R_1$ and $R_2$ remain near unity even though the time-accumulated phase density is nearly uniform. In the uncoupled pinning regime, many oscillators settle near two approximately antipodal sectors and $R_2\gg R_1$, with a residual population of forward drifters. The backward-slip column is different: both $R_1(t)$ and $R_2(t)$ undergo large repeated excursions, and among the four representative parameter sets $\delta R_1$ from Eq.~\eqref{eq:deltaR1} is largest precisely where $f_-$ is appreciable. In coherent pinning, the population settles into one stationary polar cluster, $R_1$ is large, $D\simeq0$, and $\delta R_1$ and $f_-$ vanish.

\begin{figure*}[!t]
\centering
\includegraphics[width=0.88\textwidth,height=0.88\textheight,keepaspectratio]{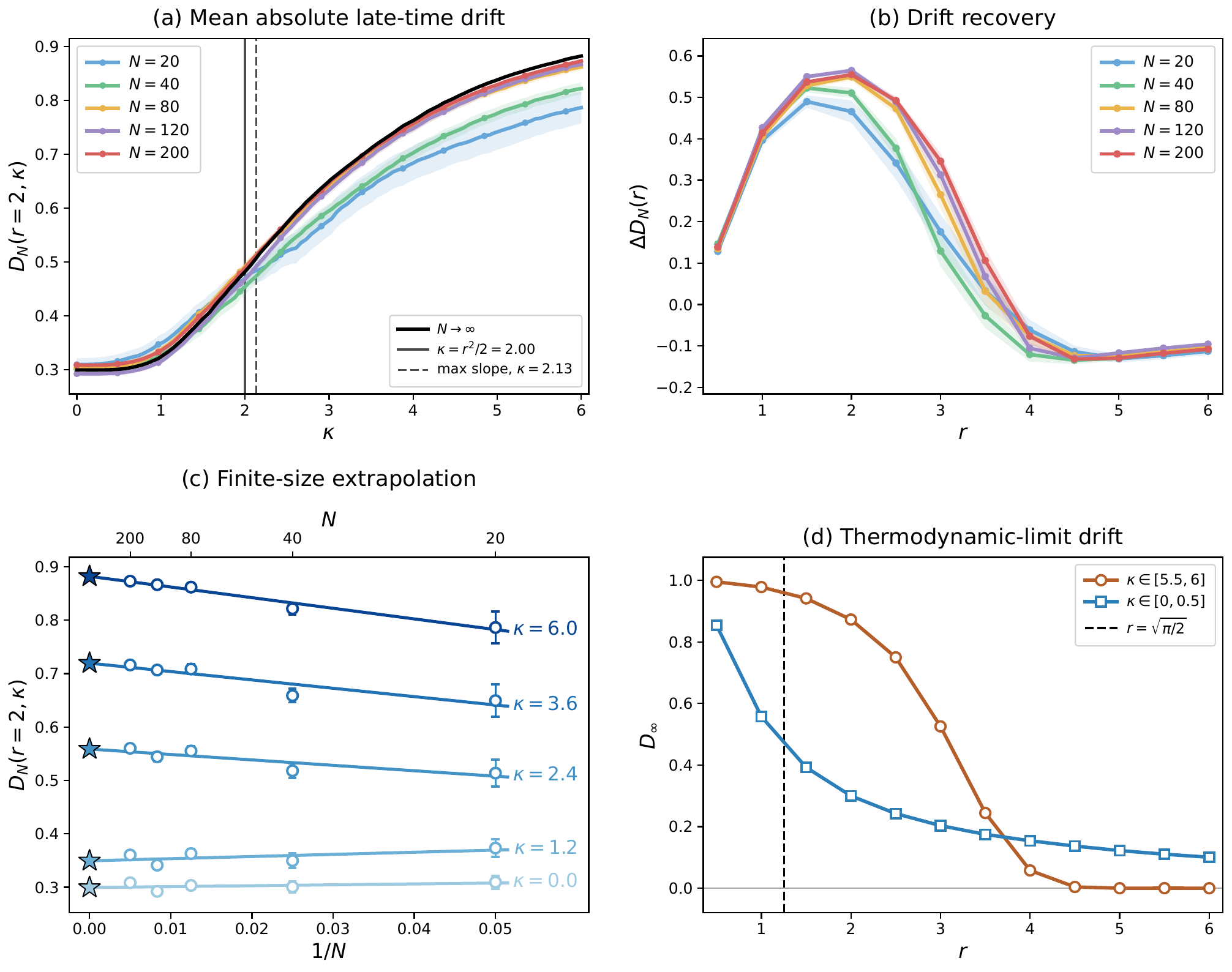}
\caption{Finite-size dependence and thermodynamic-limit extrapolation for $N=20,40,80,120,200$. (a) Mean absolute late-time drift $D_N(r=2,\kappa)$; colored curves show finite-$N$ results and the black curve shows the extrapolated $N\to\infty$ limit. The vertical lines mark $\kappa=r^2/2=2.00$ and the maximum-slope point at $\kappa=2.13$. (b) Drift recovery $\Delta D_N(r)$ between the weak- and strong-coupling windows of Eq.~\eqref{eq:coupling-windows}. (c) Representative linear extrapolations in $1/N$ at $r=2$; stars mark the $N\to\infty$ intercepts. (d) Extrapolated weak- and strong-coupling drift $D_\infty(r)$; the dashed line marks $r=\sqrt{\pi/2}$.}
\label{fig:finite-size}
\end{figure*}

\subsection{Finite-size dependence and the thermodynamic limit}

We finally examine how the drift response changes with system size. We write $D_N$ for the observable in Eq.~\eqref{eq:drift-observable} evaluated at finite $N$, and use the five sizes $N=20, 40, 80, 120, 200$. Figure~\ref{fig:finite-size} summarizes the finite-size results and the thermodynamic-limit extrapolation in four panels.

At fixed $r=2$, panel (a) shows $D_N(r=2,\kappa)$ across the coupling range. The low-coupling values depend only weakly on $N$, whereas finite-size separation becomes more visible as coupling restores drift. To quantify the effect of coupling across disorder strength, we define

\begin{equation}\label{eq:deltaD}
\Delta D_N(r)=D_N^{\mathrm{high}}(r)-D_N^{\mathrm{low}}(r),
\end{equation}

where

\begin{equation}\label{eq:coupling-windows}
\begin{aligned}
D_N^{\mathrm{low}}(r)&=\langle D_N(r,\kappa)\rangle_{\kappa\in[0,0.5]},\\
D_N^{\mathrm{high}}(r)&=\langle D_N(r,\kappa)\rangle_{\kappa\in[5.5,6]}.
\end{aligned}
\end{equation}

Panel (b) shows $\Delta D_N(r)$. As $r$ grows from small values, the response first increases because strong coupling restores motion that is suppressed in the weak-coupling window. At larger $r$, $\Delta D_N$ decreases and becomes negative because the high-coupling window enters the coherently pinned regime while a residual locally drifting fraction survives at weak coupling.

To estimate the thermodynamic limit from the same simulations, we fit the finite-size data at each $(r,\kappa)$ using

\begin{equation}\label{eq:finite-size-ansatz}
D_N(r,\kappa)=D_\infty(r,\kappa)+\frac{C(r,\kappa)}{N}+\cdots.
\end{equation}

Panel (c) displays representative fits at $r=2$ for the cuts $\kappa=0,1.2,2.4,3.6,6$. The circles and error bars are the finite-$N$ trial means and their standard errors, the straight lines are unweighted linear least-squares fits of Eq.~\eqref{eq:finite-size-ansatz}, and the stars at $1/N=0$ are the extrapolated intercepts.

The black curve in panel (a) is obtained by repeating this $1/N$ extrapolation at every sampled $\kappa$, giving $D_\infty(r=2,\kappa)$ across the full coupling range. A linear interpolation of the extrapolated curve is used to locate the maximum slope, which occurs at $\kappa\simeq2.13$, marked by the dashed line. We use this only as a measure of the drift-recovery crossover. The solid line at $\kappa=r^2/2=2.00$ shows the upper scale
from Eq.~\eqref{eq:coherent-upper-scale}; the strong-coupling approximation
is not quantitatively controlled at $r=2$, so it is included only for
qualitative comparison.

Panel (d) shows the thermodynamic-limit drift in the weak- and strong-coupling windows. At weak coupling, the extrapolated drift decreases with $r$ but remains finite over the sampled range, consistent with low-$\kappa$, large-$r$ behavior. At strong coupling, it falls rapidly to zero as disorder increases, supporting the persistence of coherent pinning in the thermodynamic limit. The dashed line marks the necessary condition for collective pinning $r=\sqrt{\pi/2}$ from Eq.~\eqref{eq:necessary-r}.

\section{Discussion and outlook}

We have studied a fully connected active rotator network in which all oscillators share the same positive intrinsic drive while the local feedback amplitudes are drawn from a zero-mean Gaussian distribution. Even without intrinsic-frequency disorder, this local feedback heterogeneity creates a nontrivial competition between individual pinning, collective alignment, and drift. Weak coupling leaves much of the dynamics close to the behavior of uncoupled active rotators, while stronger coupling can either restore a coherent forward drift or, at sufficiently large feedback disorder, organize the population into a coherently pinned state.

The symmetry $\theta_i\to\theta_i+\pi$, $a_i\to-a_i$ implies a pair of equivalent collective states. A given realization may select either branch, while averaging over both restores the ensemble symmetry. The same symmetry also makes it useful to consider both the polar order $R_1$ and the nematic order $R_2$ of the system. At large $r$ and weak coupling, the drift-suppressed state is mainly antipodal and nematically ordered, with small $R_1$, large $R_2$, and some residual drift. At strong coupling, the pinned state is instead polar, with large $R_1$ and nearly zero drift. Thus regions with similarly small $D$ can correspond to different types of phase organization.

The stationary mean-field theory describes the coherently pinned regime and agrees well with simulations at representative high-coupling points. Its continued coherent branch is fully pinned, and its lower existence edge lies close to the low-drift boundary seen numerically. The strong-coupling approximation gives the complementary asymptotic scale $\kappa\lesssim r^2/2$, which is distinct from the mean-field existence edge.

The negative-drift regime is inherently non-stationary. In a stationary mean field, oscillators can only rotate forward or become pinned, so sustained backward motion cannot arise from the local feedback alone. Negative drift appears when the collective field varies strongly in time and repeatedly pushes some oscillators through the pinning threshold, producing backward phase slips. It therefore reflects a competition between the time-dependent coupling field and the local feedback.

The finite-size results show that these trends are not restricted to a single system size. Weak-coupling drift changes only weakly with $N$, while finite-size corrections are more visible during coupling-induced drift recovery. The $1/N$ extrapolation preserves the recovery crossover and shows strong suppression of the high-coupling drift at large disorder, where the extrapolated values approach zero within the resolution of the extrapolation. This behavior supports the persistence of coherent pinning toward the thermodynamic limit, although the system sizes studied do not justify a sharper claim about critical scaling.

Several extensions follow naturally from the present work. Here, we use fully connected coupling; random or spatially embedded networks may alter both synchronization and the organization of feedback--phase correlations \cite{lopes2016,rodrigues2016}. The distribution of feedback amplitudes could also be generalized beyond a symmetric Gaussian, for example by introducing skewness, bounded disorder, or correlations with coupling strength or intrinsic frequency. Such correlations can qualitatively reorganize collective behavior in related active-rotator and Kuramoto-type systems \cite{sonnenschein2014}. Analytical reductions beyond the stationary mean-field closure would be useful for the time-dependent collective regimes. One possible route is an Ott--Antonsen formulation of the time-dependent continuum dynamics, although the Gaussian feedback distribution does not yield the simple closure available for Lorentzian disorder \cite{ott2008}. Finite-size effects and quenched heterogeneity in active rotator populations also remain natural points of comparison with earlier studies \cite{komin2010,klinshov2019}.

The model considered here is minimal, but active rotator and Kuramoto-type descriptions are widely used as reduced models for oscillatory and excitable units in physical, biological, and nonlinear dynamical systems \cite{acebron2005,shinomoto1986,sakaguchi1986,sakaguchi1988,rodrigues2016,winfree1967,strogatz2000}. In this broader setting, the present results show that mixed-sign feedback disorder can reshape not only whether a population drifts or pins, but also how its collective phase order is organized.

\section*{Acknowledgements}

I thank Sitabhra Sinha of the Institute of Mathematical Sciences, Chennai, for introducing me to the Kuramoto model in an earlier academic context. The present work was carried out independently. I thank Benoît Mahault and Andrea Parmeggiani of Laboratoire Charles Coulomb, Montpellier, and Hugues Chaté of CEA-Saclay, France, for useful discussions and feedback on this work. I also thank Jean-Charles Walter and Linda Delimi of Laboratoire Charles Coulomb, Montpellier, and Tapati Dutta of St. Xavier's College, Kolkata, for their encouragement and support.

\section*{Data availability}

The simulation code, processed data, and generated figures associated with this work are publicly available on GitHub at \url{https://github.com/arpand2004/active-rotator-drift-pinning-mixed-feedback} and archived on Zenodo at \url{https://doi.org/10.5281/zenodo.22690705}.

\section*{Funding and conflict of interest statement}

This work was carried out independently and did not receive any external funding. The author declares no competing financial or non-financial interests.

\FloatBarrier


\begin{thebibliography}{99}

\bibitem{acebron2005}
J. A. Acebr\'on, L. L. Bonilla, C. J. P\'erez Vicente, F. Ritort, and R. Spigler, ``The Kuramoto model: A simple paradigm for synchronization phenomena,'' \emph{Reviews of Modern Physics} \textbf{77}, 137--185 (2005). \url{https://doi.org/10.1103/RevModPhys.77.137}

\bibitem{shinomoto1986}
S. Shinomoto and Y. Kuramoto, ``Phase Transitions in Active Rotator Systems,'' \emph{Progress of Theoretical Physics} \textbf{75}, 1105--1110 (1986). \url{https://doi.org/10.1143/PTP.75.1105}

\bibitem{sakaguchi1986}
H. Sakaguchi and Y. Kuramoto, ``A Soluble Active Rotator Model Showing Phase Transitions via Mutual Entrainment,'' \emph{Progress of Theoretical Physics} \textbf{76}, 576--581 (1986). \url{https://doi.org/10.1143/PTP.76.576}

\bibitem{sakaguchi1988}
H. Sakaguchi, S. Shinomoto, and Y. Kuramoto, ``Phase transitions and their bifurcation analysis in a large population of active rotators with mean-field coupling,'' \emph{Progress of Theoretical Physics} \textbf{79}, 600--607 (1988). \url{https://doi.org/10.1143/PTP.79.600}

\bibitem{strogatz1989}
S. H. Strogatz, C. M. Marcus, R. M. Westervelt, and R. E. Mirollo, ``Collective dynamics of coupled oscillators with random pinning,'' \emph{Physica D: Nonlinear Phenomena} \textbf{36}, 23--50 (1989). \url{https://doi.org/10.1016/0167-2789(89)90246-7}

\bibitem{lopes2016}
M. A. Lopes, E. M. Lopes, S. Yoon, J. F. F. Mendes, and A. V. Goltsev, ``Synchronization in the random-field Kuramoto model on complex networks,'' \emph{Physical Review E} \textbf{94}, 012308 (2016). \url{https://doi.org/10.1103/PhysRevE.94.012308}

\bibitem{komin2010}
N. Komin and R. Toral, ``Order parameter expansion and finite-size scaling study of coherent dynamics induced by quenched noise in the active rotator model,'' \textit{Physical Review E} \textbf{82}, 051127 (2010). \url{https://doi.org/10.1103/PhysRevE.82.051127}

\bibitem{klinshov2019}
V. Klinshov and I. Franovi\'c, ``Two scenarios for the onset and suppression of collective oscillations in heterogeneous populations of active rotators,'' \emph{Physical Review E} \textbf{100}, 062211 (2019). \url{https://doi.org/10.1103/PhysRevE.100.062211}

\bibitem{rodrigues2016}
F. A. Rodrigues, T. K. D. M. Peron, P. Ji, and J. Kurths, ``The Kuramoto model in complex networks,'' \emph{Physics Reports} \textbf{610}, 1--98 (2016). \url{https://doi.org/10.1016/j.physrep.2015.10.008}

\bibitem{sonnenschein2014}
B. Sonnenschein, T. K. D. M. Peron, F. A. Rodrigues, J. Kurths, and L. Schimansky-Geier, ``Cooperative behavior between oscillatory and excitable units: the peculiar role of positive coupling-frequency correlations,'' \emph{European Physical Journal B} \textbf{87}, 182 (2014). \url{https://doi.org/10.1140/epjb/e2014-50274-2}

\bibitem{ott2008}
E. Ott and T. M. Antonsen, ``Low dimensional behavior of large systems of globally coupled oscillators,'' \emph{Chaos} \textbf{18}, 037113 (2008). \url{https://doi.org/10.1063/1.2930766}

\bibitem{winfree1967}
A. T. Winfree, ``Biological rhythms and the behavior of populations of coupled oscillators,'' \emph{Journal of Theoretical Biology} \textbf{16}, 15--42 (1967). \url{https://doi.org/10.1016/0022-5193(67)90051-3}

\bibitem{strogatz2000}
S. H. Strogatz, ``From Kuramoto to Crawford: exploring the onset of synchronization in populations of coupled oscillators,'' \emph{Physica D: Nonlinear Phenomena} \textbf{143}, 1--20 (2000). \url{https://doi.org/10.1016/S0167-2789(00)00094-4}

\end{thebibliography}
\end{document}